\documentclass{svproc}

\usepackage{url}
\usepackage{graphicx}
\usepackage{multirow}

\newcommand{\be}{\begin{equation}}
\newcommand{\ee}{\end{equation}}
\newcommand{\bea}{\begin{eqnarray}}
\newcommand{\eea}{\end{eqnarray}}

\newcommand{\xv}{{\mathbf x}}
\newcommand{\vecnul}{{\mathbf 0}}
\newcommand{\bra}{\langle}
\newcommand{\ket}{\rangle}

\newcommand{\eps}{\epsilon}

\newcommand{\cC}{${\cal C}\;$}

\newcommand{\half}{\frac{1}{2}}

\begin{document}
\mainmatter              
\title{Hyperons in thermal QCD from the lattice$^\ast$}
\titlerunning{Hyperons in thermal QCD}  
%
\author{Gert Aarts\inst{1} \and
Chris Allton\inst{1}$^\dagger$ \and
Davide de Boni\inst{1} \and
Jonas Glesaaen\inst{1} \and
Simon Hands\inst{1} \and
Benjamin Jäger\inst{2} \and
Jon-Ivar Skullerud\inst{3}
}
\authorrunning{Chris Allton et al.} 
\institute{
Department of Physics, College of Science, Swansea University,
Swansea SA2 8PP, United Kingdom
\and
CP$^3$-Origins \& Danish IAS, Department of Mathematics and Computer
Science, University of Southern Denmark, 5230 Odense M, Denmark
\and
Department of Theoretical Physics, National University of
Ireland Maynooth, Maynooth, County Kildare, Ireland \\
$\dagger$Speaker \email{c.allton@swansea.ac.uk}\\
}

\maketitle              



\begin{abstract}
We study the spectrum of light baryons and hyperons as a function of
temperature using lattice gauge theory methods.  We find that masses
of positive parity states are temperature independent, within errors, in
the hadronic phase.  The negative parity states decrease in mass as
the temperature increases.  Above the deconfining temperature, lattice
correlators and spectral functions show a degeneracy between parity
sectors, i.e. parity doubling.  We apply our findings to an in-medium
Hadron Resonance Gas model. The techniques used in this study include
direct analysis of the hadronic correlation functions, conventional
fitting procedures, and the Maximum Entropy Method.  
\end{abstract}



\section{Introduction}

Symmetries play a crucial role in the Standard Model and are
especially significant in the QCD transition from the confining,
chirally broken, hadronic phase to the deconfined, chirally symmetric,
plasma phase.  Chiral symmetry restoration in the meson sector at
finite temperature has been studied extensively, but there have been
very few studies in baryons \cite{DeTar:1987ar,Pushkina:2004wa,Datta:2012fz}.

In the baryonic case, a combination of unbroken chiral symmetry and
parity leads to parity doubling, i.e. a degeneracy between positive
and negative parity states.  Hence it is expected that in the chirally
symmetric phase, baryonic channels related by parity will display
approximate degeneracy.  This work tests this conjecture by studying octet
and decuplet light baryons and hyperons as the temperature varies
using our FASTSUM Collaboration's anisotropic, 2+1 flavour lattices
\cite{Aarts:2014nba}.

We find evidence of parity doubling above the deconfining temperature,
$T_c$,\footnote{$T_c$ is not uniquely defined. 
 Here it's defined from the renormalised Polyakov loop
  \cite{Aarts:2014nba}.\newline\hspace*{-4mm}
$\ast$ Invited plenary talk presented at the Strangeness in Quark
Matter Conference (SQM 2019), Bari, Italy, 10-15 June 2019.}
with this degeneracy being most pronounced
for the baryons with the smallest strange quark content.  In the
hadronic phase we find that the negative parity masses decrease as $T
\rightarrow T_c$ while the corresponding positive parity masses
remain constant within the uncertainty. We use these
temperature-dependent masses to define an in-medium Hadron Resonance
Gas, deriving results for susceptibilities and partial pressures.

The work presented here is detailed more fully in
\cite{Aarts:2015mma,Aarts:2017rrl,Aarts:2018glk}.



\section{Parity in Baryons}

We use the standard interpolation operator for a nucleon
\[
O_N(\xv,\tau) = \eps_{abc}
u_a(\xv,\tau) \left[ u_b^T(\xv,\tau) {\cal C} \gamma_5 d_c(\xv,\tau)
\right],
\nonumber
\]
where $u, d$ are the quark fields, $a,b,c$ are colour indices, other
indices are suppressed and \cC denotes the charge conjugation matrix.
Similar operators are used for octet and decuplet cases, see
\cite{Aarts:2017rrl}.
Parity operators for the positive and negative parity
channels are defined as
$O_{N_\pm}(\xv,\tau) = P_\pm O_N(\xv,\tau),$
where $P_\pm=\half(1\pm\gamma_4).$
%
We study the usual Euclidean correlators of these operators, summed
over the Dirac indices and projected to zero momentum,
\[
G_\pm(\tau) = \int d^3x\,
\left\bra O_{N_\pm}(\xv,\tau) \overline{O}_{N_\pm}(\vecnul,0) \right\ket.
\]
From the properties of Euclidean time reflection, it follows that the
forward (backward) propagation of $G_+(\tau)$ corresponds to the
positive (negative) parity channel (see e.g. \cite{Aarts:2015mma}).
Hence both parities are obtained from one correlator.



\section{Lattice parameters}

Our FASTSUM collaboration specialises in using a fixed-scale approach
on an\-iso\-tropic lattices where the temporal lattice spacing $a_\tau$ is
smaller than the spatial one $a_s$.  We use 2+1 quark flavours,
where the strange quark mass has its physical value, but the two
lightest flavours are heavier than in nature resulting in a pion mass
of $M_\pi = 392(4)$ MeV.  
The inverse temporal lattice spacing is $a_\tau^{-1} = 5.63(4) $GeV with
$a_s/a_\tau \approx 3.5$, and our spatial lattice volume is $24^3$.  We use a
variety of temporal extents, $N_\tau$, with the corresponding
temperatures $T = 1/(a_\tau N_\tau)$ shown in Table \ref{tab:lat},
spanning both phases.

\begin{table}
\begin{center}
\begin{tabular}{l|rrrrrrrr}
\hline
$N_\tau$     &        128 &   40 &   36 &   32 &   28 &   24 &   20 &   16 \\
$T/T_c\;\;$  &$\;\;$ 0.24 & 0.76 & 0.84 & 0.95 & 1.09 & 1.27 & 1.52 & 1.90 \\
$T$ [MeV]    &         44 &  141 &  156 &  176 &  201 &  235 &  281 & 352 \\
\hline
\end{tabular}
\end{center}
\caption{Lattice parameters and temperatures studied. The ensemble at
  the lowest temperature was provided by the HadSpec Collaboration
  \cite{Edwards:2008ja}
\label{tab:lat}}
\end{table}



\section{Results}

\begin{figure}
\centering
\includegraphics[width=0.7\textwidth]{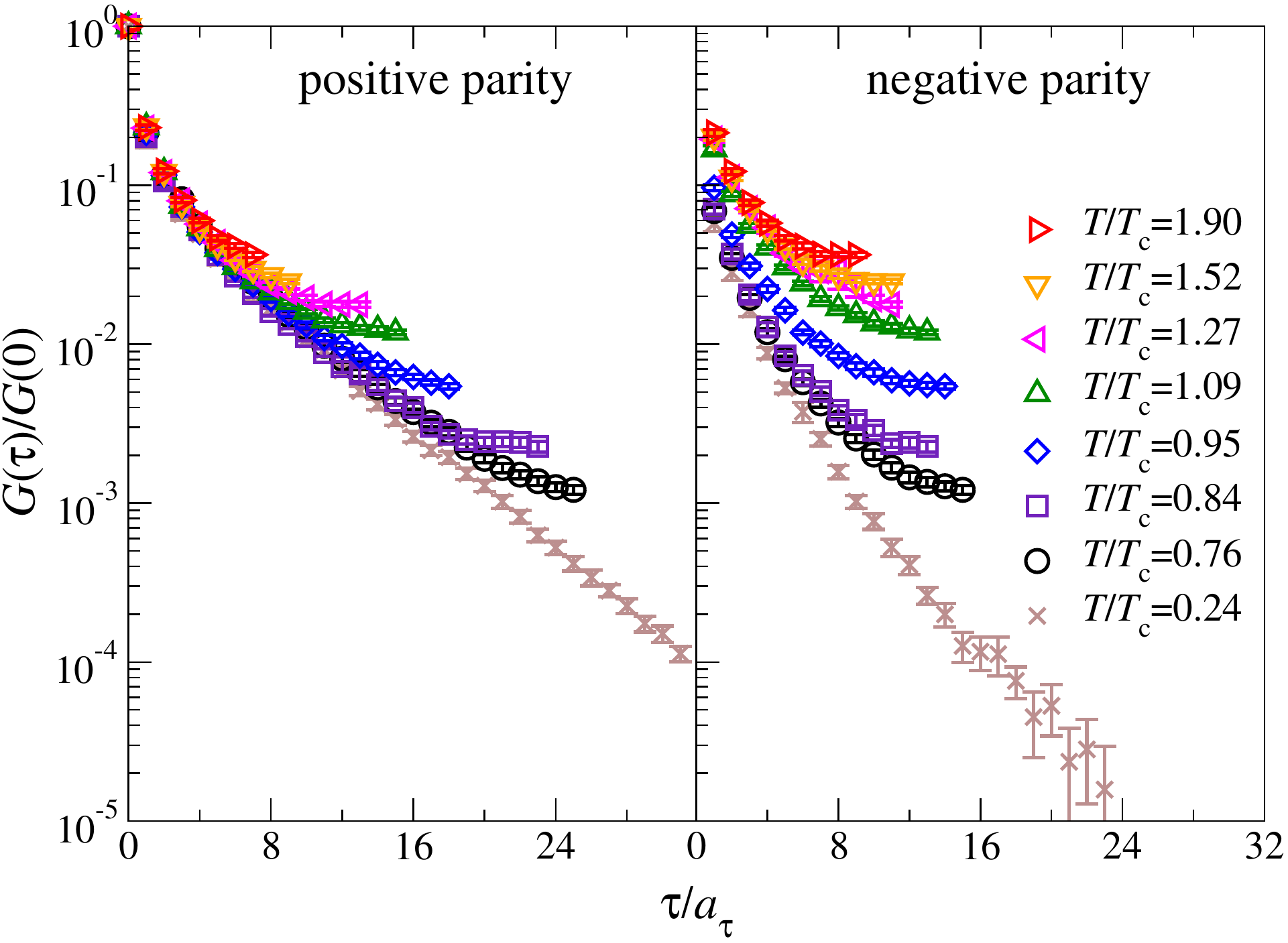}
\caption{Temperature dependent correlators for the +ve and
  -ve parity nucleon channels.
}
\label{fig:corfns}
\end{figure}

The nucleon correlation functions are shown in Fig.~\ref{fig:corfns} for both
parity states. Other channels have similar behaviour.
The backward movers (from the second half of the temporal range) have
been reflected in $\tau$ to enable direct comparison with their
positive counterparts.


\begin{figure}
\centering
\includegraphics[width=0.49\textwidth]{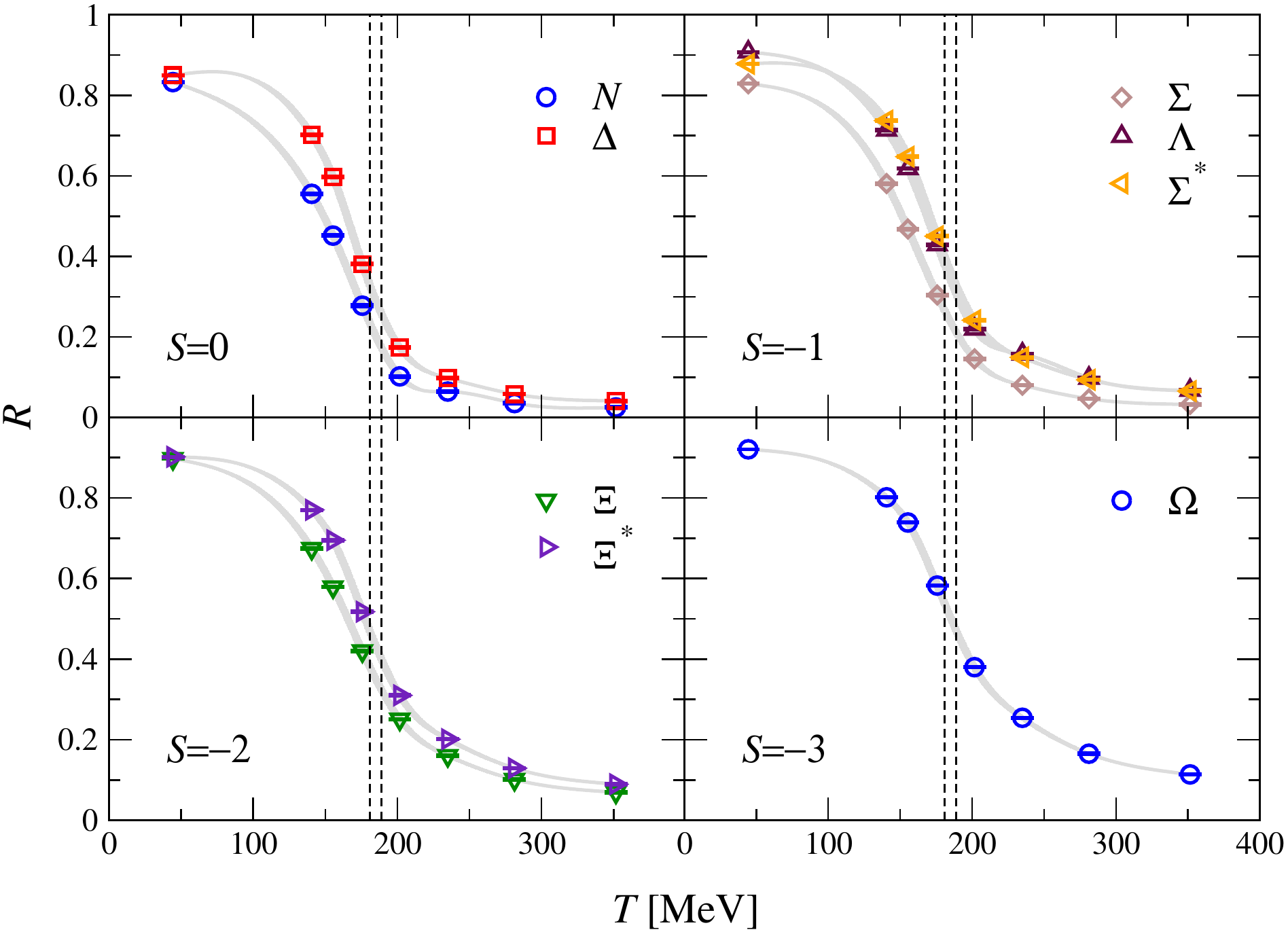}
\includegraphics[width=0.49\textwidth]{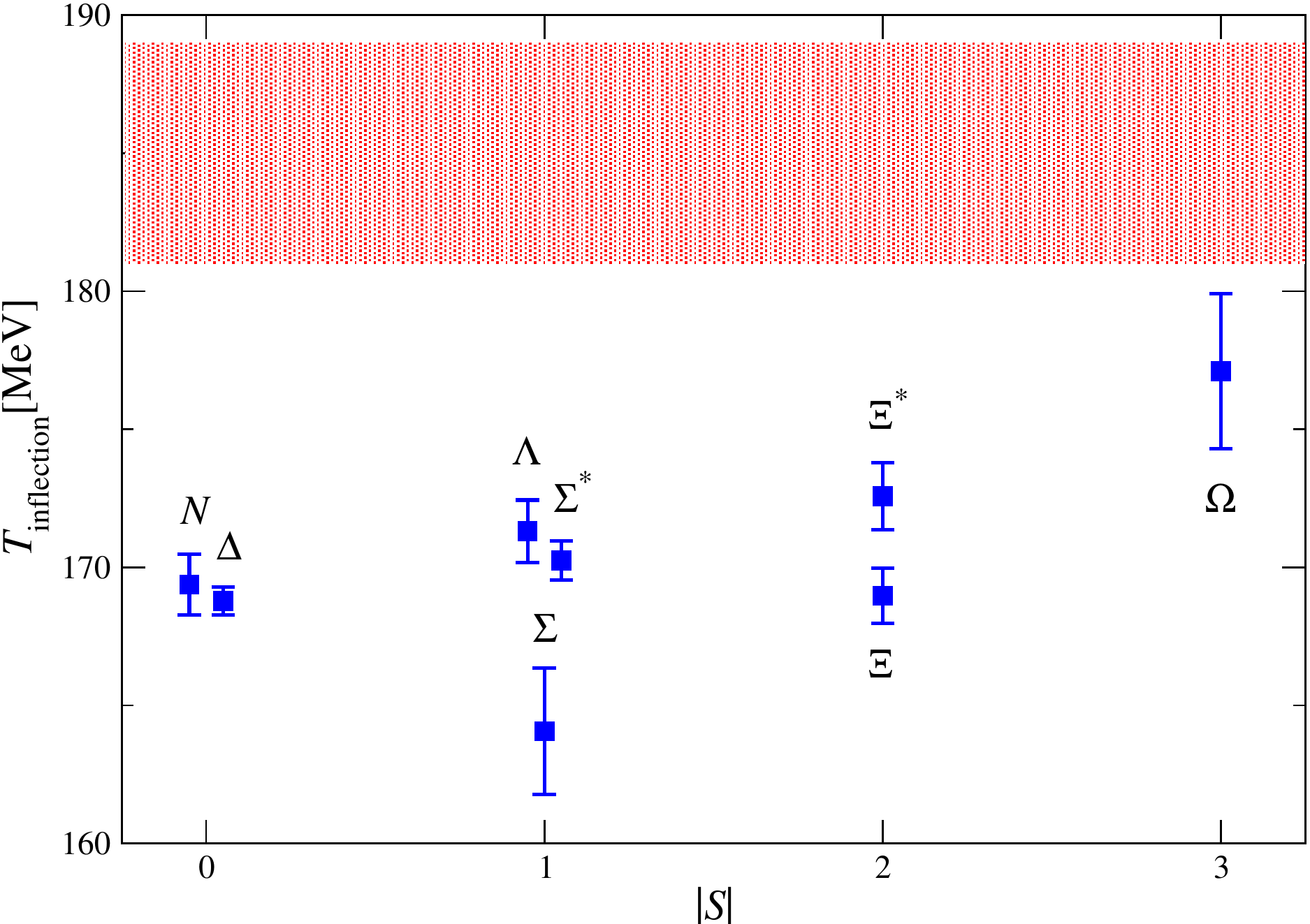}
\caption{
{\bf (Left)}
The ratio $R$ for various strangeness sectors.
{\bf (Right)}
Inflection point temperature
of $R$ with $T_c$ from the Polyakov loop shown in the band
\cite{Aarts:2014nba}.}
\label{fig:R}
\label{fig:Tinflection}
\end{figure}


We define the ratio
$R(\tau) = [G_+(\tau) - G_+(1/T - \tau)]/[G_+(\tau) + G_+(1/T - \tau)]$
which measures the time reflection (a)symmetry of $G_+(\tau)$ and therefore
the presence (absence) of parity doubling.
$R(\tau)\sim 1$ signifies non-degenerate parity states, and 
$R(\tau)\sim 0$ parity doubling.
It is convenient to average $R(\tau)$ over $\tau$ obtaining
$R$, a quasi order parameter,
and this is plotted in Fig.~\ref{fig:R} (left).
$R$ clearly approaches zero around $T_c$
indicating parity doubling.
This effect is strongest for baryons with the smallest strangeness
content, as expected due to the residual chiral symmetry breaking effects
from the strange quark.

We obtain a measure of the transition temperature for each channel
using the point of inflection of $R$
and plot these in Fig.~\ref{fig:R} (right) together with the $T_c$
value obtained from the Polyakov loop \cite{Aarts:2014nba}.

\begin{figure}
\centering
\includegraphics[width=0.49\textwidth]{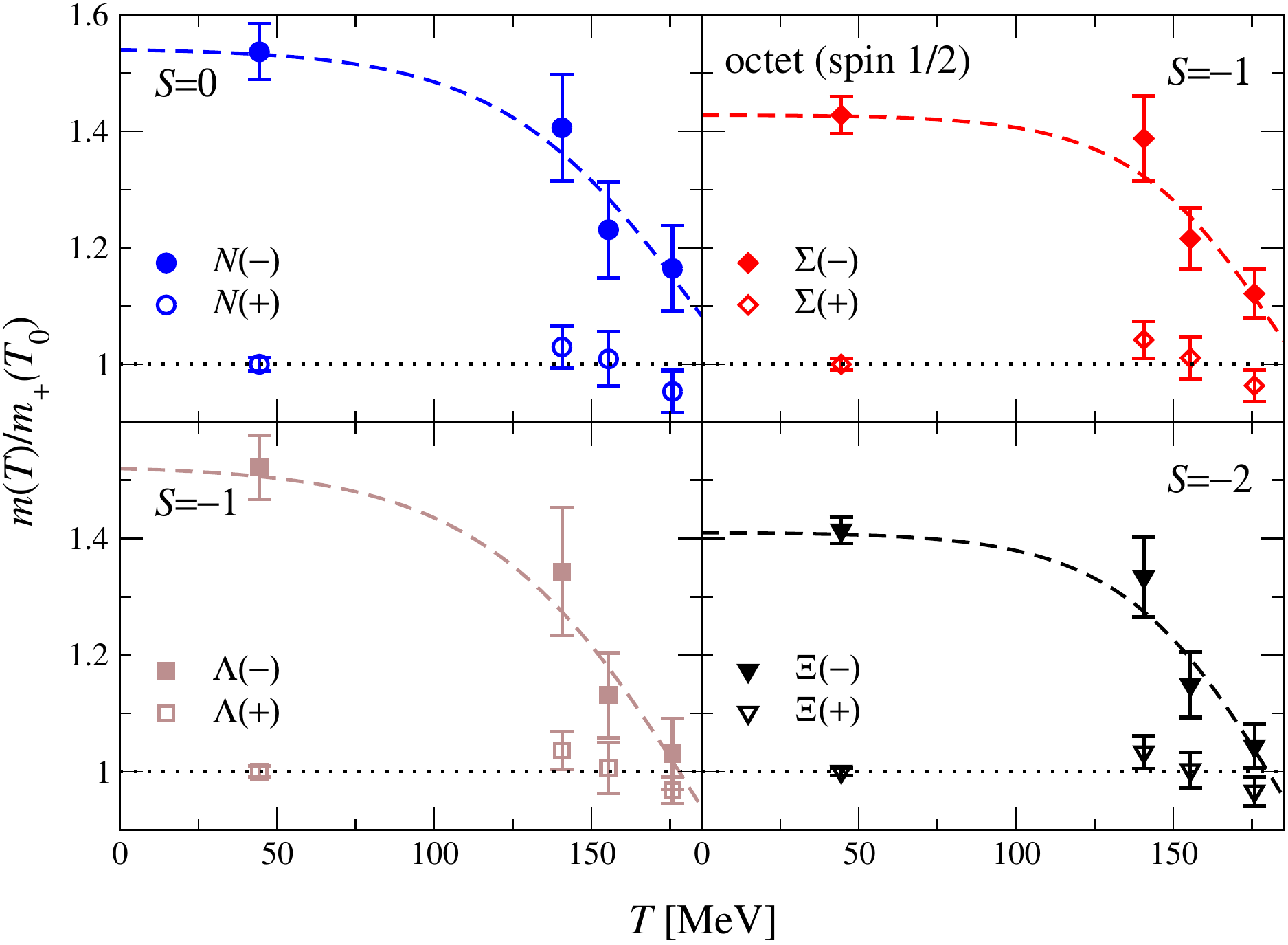}
\includegraphics[width=0.49\textwidth]{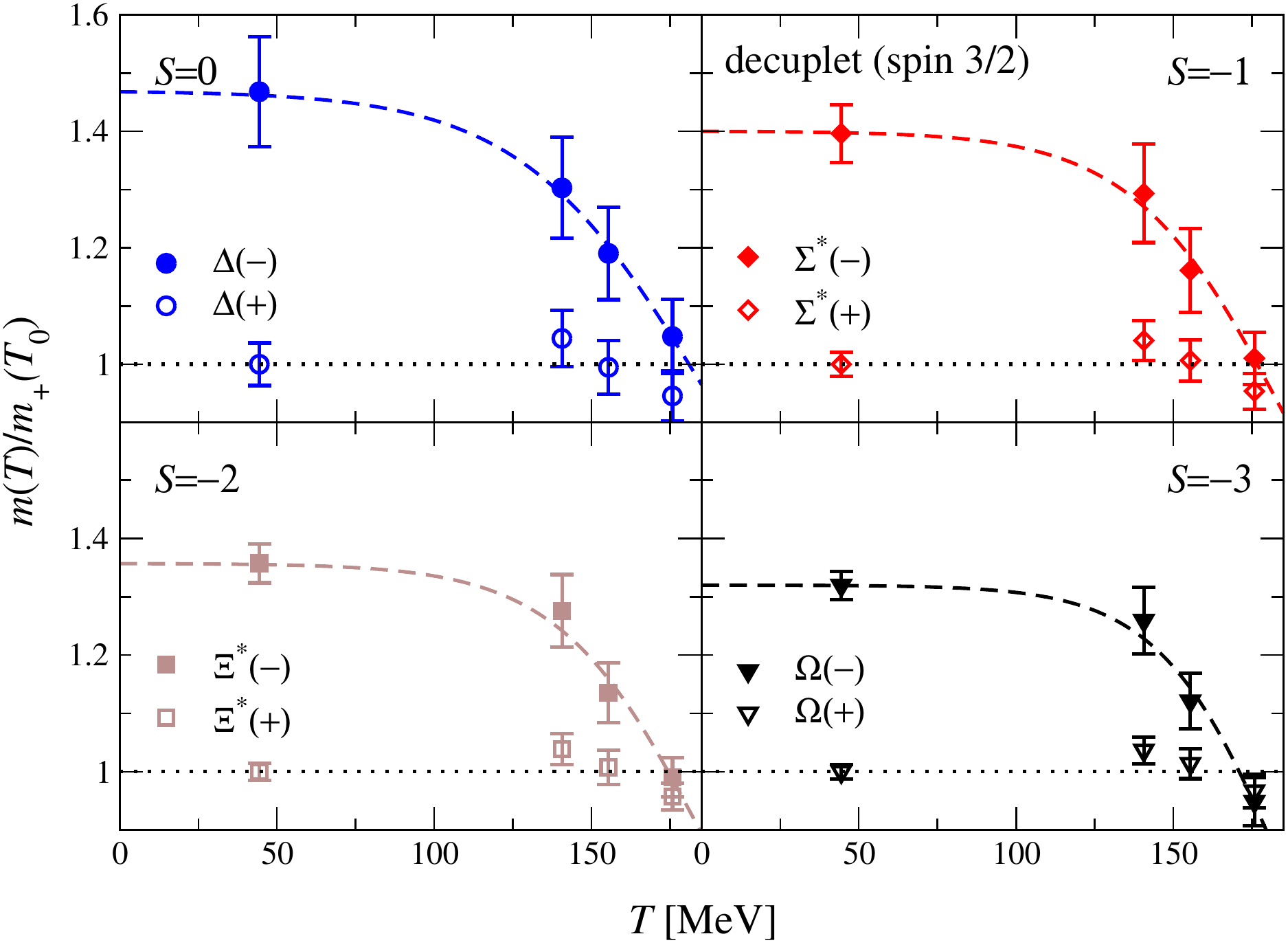}
\caption{Temperature dependent masses of the octet and decuplet
  baryons for both parity sectors.
\label{fig:MvsT}}
\end{figure}

The ground state masses are extracted from the correlators in the
hadronic phase using conventional exponential fits with the results
displayed in Fig.~\ref{fig:MvsT}.
We see that the $T=0$ masses are heavier than the experimental values
-- this reflects the non-physical value of the two lightest dynamical
quarks in our simulation. The positive parity states' masses appear to be
$T$-independent, whereas the negative states' masses decrease as
$T\rightarrow T_c$.


\begin{figure}
\centering
\includegraphics[width=0.51\textwidth]{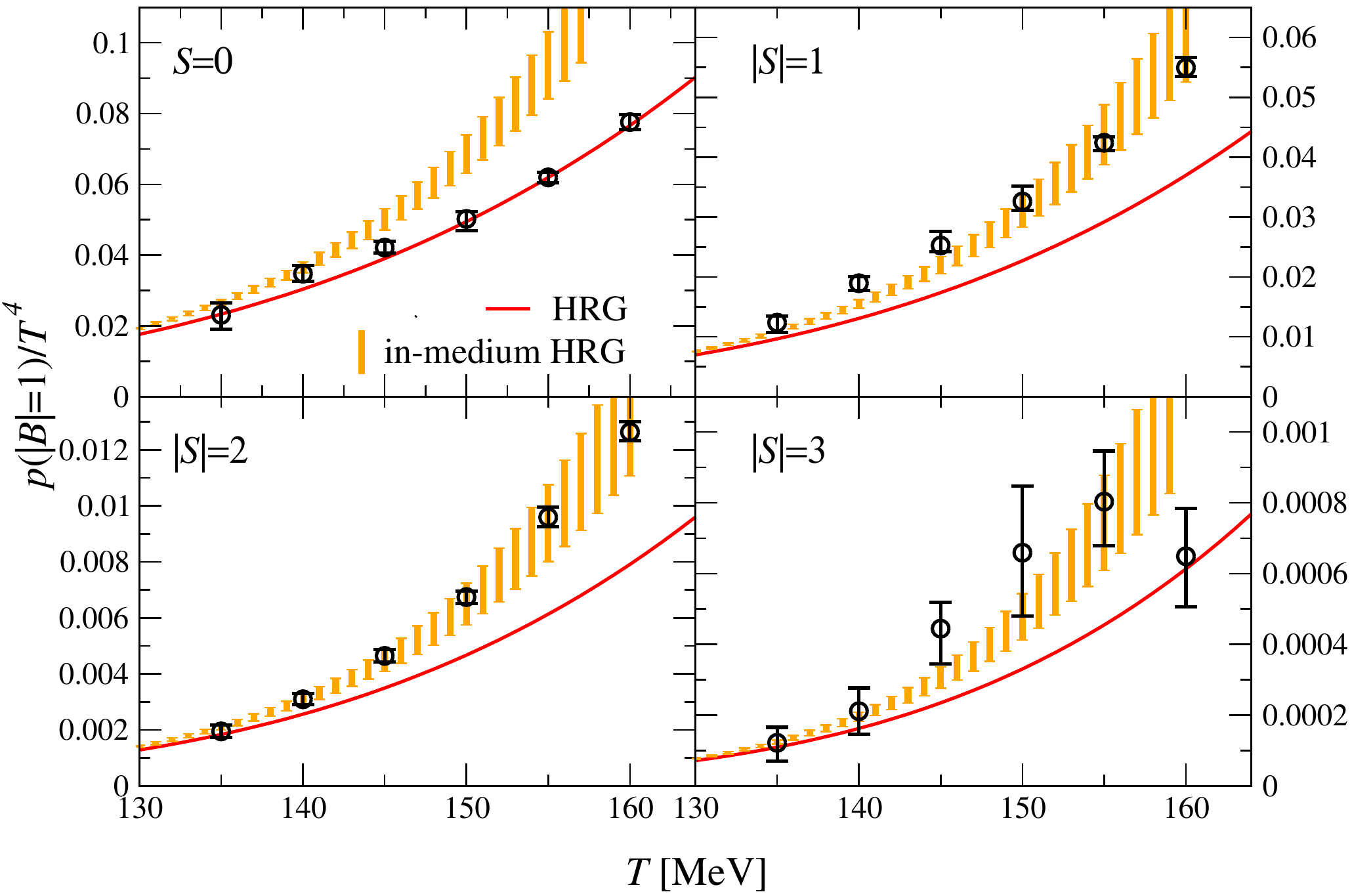}
\includegraphics[width=0.48\textwidth]{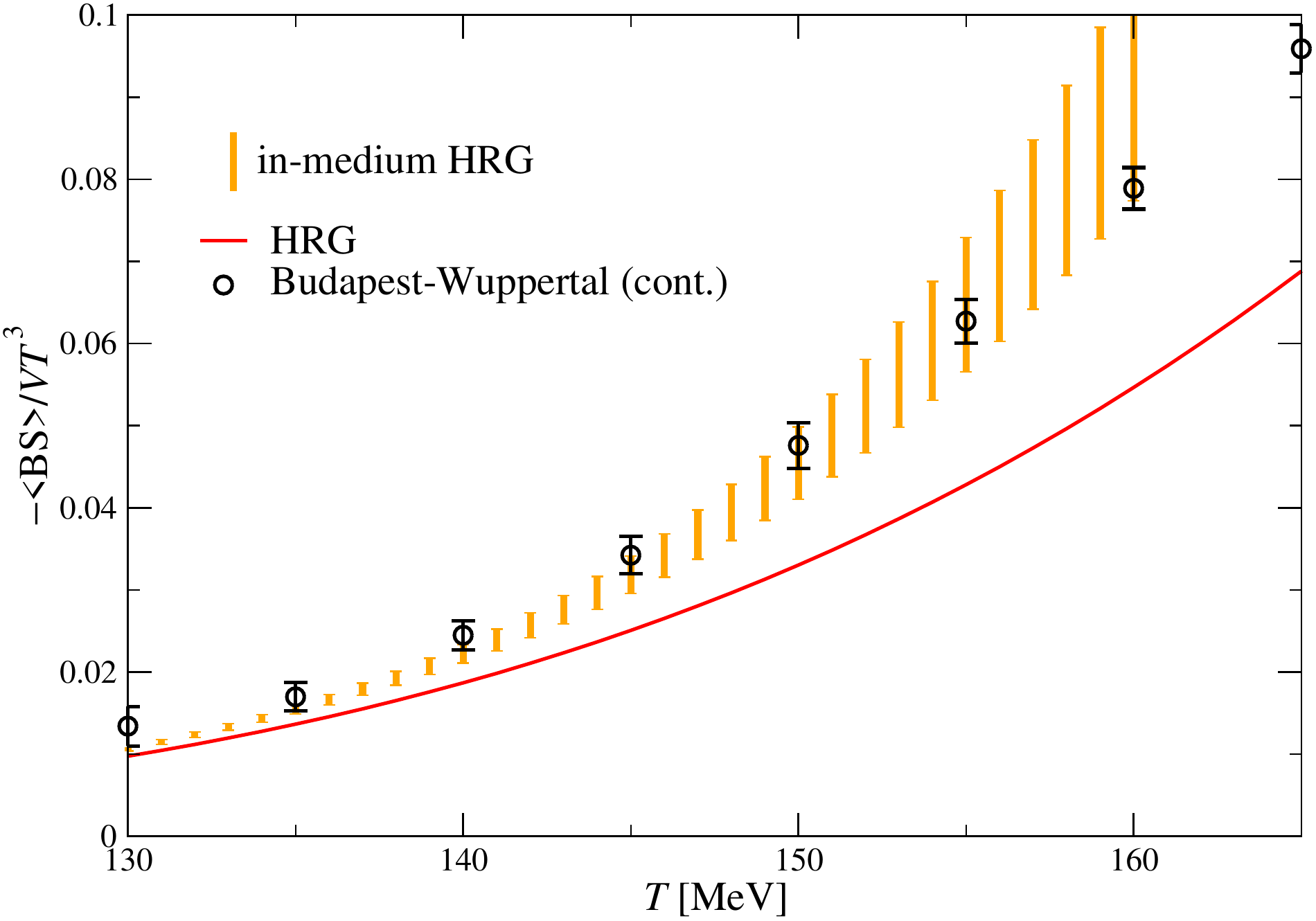}
\caption{In-medium Hadron Resonance Gas results. (Left) partial
  pressures for each strangeness sector, and (Right) the
  susceptibility $\bra BS\, \ket$.}
\label{fig:HRG}
\end{figure}

By using these temperature-dependent masses in the Hadron Resonance
Gas (HRG) model, we obtain ``in-medium HRG'' predictions for the
partial pressures and susceptibility $\chi_{BS} = \frac{1}{VT}\bra
BS\, \ket$, where $B$ and $S$ are the baryon and strangeness number,
see Fig.~\ref{fig:HRG}.  For $\chi_{BS}$ and the strange partial
pressure sectors, this in-medium HRG gives better agreement with
independently obtained lattice results \cite{Borsanyi:2011sw}.

\begin{figure}
\centering
\includegraphics[width=0.49\textwidth]{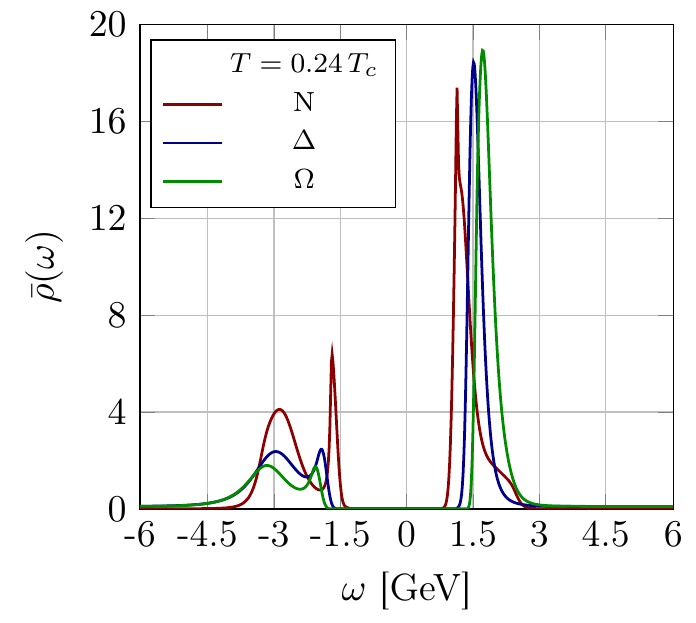}
\includegraphics[width=0.49\textwidth]{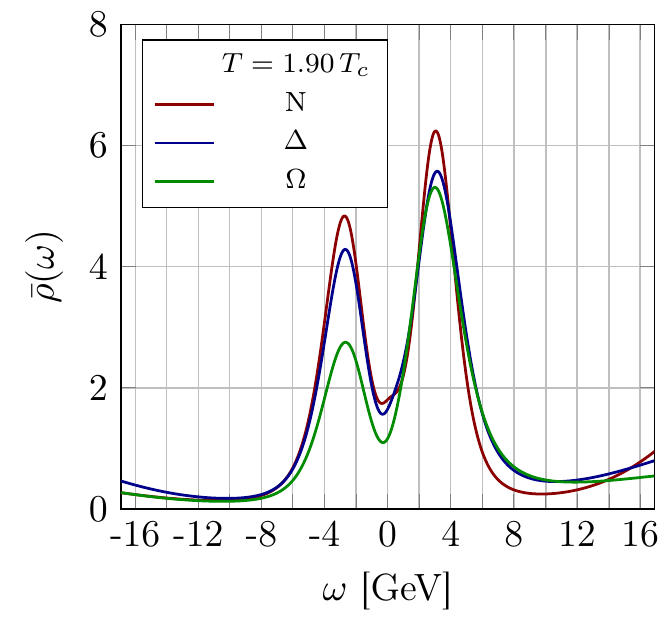}
\caption{Spectral function for indicative temperatures in the (left)
  hadronic and (right) plasma phases for three baryon channels.
The details of our Maximum Entropy Method procedure, including choices
of default model are detailed in \cite{Aarts:2017rrl}.
}
\label{fig:mem}
\end{figure}

To interpret hadronic correlation functions in the plasma phase, where
exponential fits no longer work and we assume there are no bound
states, we introduce spectral functions \cite{Aarts:2017rrl},
\[
G_\pm(\tau) = \int_{-\infty}^{\infty} \; \frac{d\omega}{2\pi} \;
 K(\tau,\omega) \; \rho_\pm(\omega),
\;\;\;\;{\rm where}\;\;
K(\tau,\omega) = \frac{e^{-\omega \tau}}{1+e^{-\omega/T}}.
\]
We use the Maximum Entropy Method to solve the above inverse problem
for $\rho(\omega)$, noting that the positive (negative) parity states
appear for $\omega>0$ $(\omega<0)$.
Spectral function results for a selection of channels are shown in
Fig.~\ref{fig:mem} for a representative temperature in each phase.
We see clear ground states in both parity sectors in the hadronic
phase, and signs of parity doubling in the plasma phase, particularly
for the non-strange baryons.


\section{Conclusion}

This work uses lattice simulations of 2+1 flavour QCD to analyse
the parity states in the baryonic spectrum as the temperature is varied
from essentially zero to around 350 MeV.
Our FASTUM collaboration's lattices are anisotropic which increases the
sampling rate in the temporal direction, thereby enhancing the
accuracy of our results.
We employ a variety of approaches in this work: using the correlation
functions directly, conventional fits to exponentials, and extracting
the spectral functions using the Maximum Entropy Method.

In the hadronic phase, these methods indicate that the positive parity
ground state masses are temperature independent (within errors) and
that the negative parity states' masses decrease as $T$ increases
until becoming essentially degenerate with its positive parity partner
at or close to $T_c$. This ``parity-doubling'' is particularly evident
for baryons with the smallest strange content, and this is to be
expected due to the strange quark's chiral symmetry breaking effects.
This pattern is observed for both octet and decuplet states.

We plan to use both lighter quarks and lattices with finer (temporal)
spacings to test the systematics of our approach in our next
generation ensembles.



\section*{Acknowledgements}

We acknowledge PRACE for awarding us access to Marconi at CINECA,
Italy.
This work used the STFC DiRAC Blue Gene Q system at the University of
Edinburgh, U.K.
We have been supported by the STFC grant ST/P00055X/1,
and the Swansea Academy for Advanced Computing.




\begin{thebibliography}{6}


\bibitem{DeTar:1987ar}
  C.~E.~DeTar and J.~B.~Kogut,
  Phys.\ Rev.\ Lett.\  {\bf 59} (1987) 399;
  Phys.\ Rev.\ D {\bf 36} (1987) 2828.

\bibitem{Pushkina:2004wa}
  I.~Pushkina {\it et al.}  [QCD-TARO Collaboration],
  Phys.\ Lett.\ B {\bf 609} (2005) 265
  [hep-lat/0410017].

\bibitem{Datta:2012fz}
  S.~Datta, S.~Gupta, M.~Padmanath, J.~Maiti and N.~Mathur,
  JHEP {\bf 1302} (2013) 145
  [arXiv:1212.2927 [hep-lat]].

\bibitem{Aarts:2014nba}
  G.~Aarts, C.~Allton, A.~Amato, P.~Giudice, S.~Hands and
  J.~I.~Skullerud,
  JHEP {\bf 1502} (2015) 186
  doi:10.1007/JHEP02(2015)186
  [arXiv:1412.6411 [hep-lat]].

\bibitem{Aarts:2015mma}
  G.~Aarts, C.~Allton, S.~Hands, B.~Jäger, C.~Praki and
  J.~I.~Skullerud,
  Phys.\ Rev.\ D {\bf 92} (2015) no.1,  014503
  [arXiv:1502.03603 [hep-lat]].

\bibitem{Aarts:2017rrl}
  G.~Aarts, C.~Allton, D.~De Boni, S.~Hands, B.~Jäger, C.~Praki and
  J.~I.~Skullerud,
  JHEP {\bf 1706} (2017) 034
  [arXiv:1703.09246 [hep-lat]].

\bibitem{Aarts:2018glk}
  G.~Aarts, C.~Allton, D.~De Boni and B.~Jäger,
  Phys.\ Rev.\ D {\bf 99} (2019) no.7,  074503
  [arXiv:1812.07393 [hep-lat]].

\bibitem{Edwards:2008ja}
  R.~G.~Edwards, B.~Joo and H.~W.~Lin,
  Phys.\ Rev.\ D {\bf 78} (2008) 054501
  doi:10.1103/PhysRevD.78.054501
  [arXiv:0803.3960 [hep-lat]].

\bibitem{Tanabashi:2018oca}
  M.~Tanabashi {\it et al.} [Particle Data Group],
  Phys.\ Rev.\ D {\bf 98} (2018) no.3,  030001.


\bibitem{Borsanyi:2011sw} 
  S.~Borsanyi, Z.~Fodor, S.~D.~Katz, S.~Krieg, C.~Ratti and K.~Szabo,
  JHEP {\bf 1201}, 138 (2012)
  [arXiv:1112.4416 [hep-lat]],
%
  R.~Bellwied, S.~Borsanyi, Z.~Fodor, S.~D.~Katz, A.~Pasztor, C.~Ratti
  and K.~K.~Szabo,
  Phys.\ Rev.\ D {\bf 92} (2015) no.11,  114505
  [arXiv:1507.04627 [hep-lat]],
%
  P.~Alba {\it et al.},
  Phys.\ Rev.\ D {\bf 96} (2017) no.3,  034517
  [arXiv:1702.01113 [hep-lat]].


\end{thebibliography}
\end{document}